\documentclass[fleqn,twoside]{article}
\usepackage{espcrc2}
\usepackage{graphicx}

\newcommand{\AmS}{{\protect\the\textfont2
  A\kern-.1667em\lower.5ex\hbox{M}\kern-.125emS}}

\title{
Vortices in superconducting strips: interplay between
        surface effects and the pinning landscape}

\author{Cl\'ecio C. de Souza Silva, Leonardo R. E. Cabral, and
J. Albino Aguiar\\ \addressmark{Departamento de F\'{\i}sica, 
Universidade Federal de Pernambuco, 50670-901 Recife, PE, Brasil}}

\begin{document}

\thispagestyle{myheadings}
\markright{\small{Presented on the II Euro-Conference on Vortex Matter in
Superconductors, Crete, 15-25 September 2001}}

\begin{abstract}
Vortices in a narrow superconducting strip with a square array of pinning 
sites are studied. The interactions of vortices with other vortices and 
with external sources (applied magnetic field and transport current) are
calculated via a screened Coulomb model. The edge barrier is taken into 
account and shown to have an important role on the system dynamics.
Numerical simulations in this approach show that the field dependent magnetic
moment presents peaks corresponding to history dependent configurations of
the vortex lattice. Some effects of the edge barrier on the I-V characteristics
are also reported.
\vspace{1pc}
\end{abstract}


\maketitle

\section{Introduction}

There is a growing interest in the microscopic structure of the
vortex lattice (VL) in superconducting films with artificially fabricated 
columnar defects. In the presence of a perpendicular magnetic field and
zero current this structure depends on how strong
is the pinning force produced by the defects with respect to the internal 
elastic forces of the VL. 
This results in two possible situations: commensurate and incommensurate 
lattice, depending on the pinning landscape being strong enough to determine
the VL structure or not. Many experimental \cite{CDexp} and theoretical
\cite{CDtheory,Gilson} works have
reported maxima in the critical current $j_c$ and field dependent magnetic
moment $m(H)$ in the points of the phase diagram where a commensurate
state takes place. 

When a finite current stronger than $j_c$ is applied, an additional
ingredient governs the structural evolution of the VL. In the center of
mass (CM) frame, the vortices interact with a time dependent pinning
potential, which is determined by the CM velocity. As the driving current
is increased from $j_c$, a variety of dynamical phases is possible. But
commensurate ordered lattices are only possible at very high driving
forces, when the pinning potential is averaged in the direction of motion
becoming a static washboard potential. Recently, Carneiro \cite{Gilson}
has shown by numerical simulations that at a high enough temperature this
ordered state is reached at a finite driving force.

Frequently, these phenomena have been investigated under the assumption 
of infinite sample, i.e. no surface or geometrical effects are taken into 
account. As it is well known, surface effects can play an important and 
even a dominant role on the irreversible properties of a superconducting 
specimen.
In the critical state approach, Kuznetsov {\it et al.} \cite{Kuznetsov}
estimated the enhancement due to edge effects of hysteresis in magnetization
of thin strips with strong pinning.
The penetration of flux in finite sample are also described by the
geometric barrier model \cite{Zeldov,Brandt01,Leo}, which has shown to be
successful in describing the irreversible properties of flat high-$T_c$
single crystals \cite{Leo} at high temperatures.
All these continuum models however are not able to determine structural
details of the VL and their relevance in the macroscopic response.

In the present work, we propose a simple qualitative description of individual
vortices dynamics in superconducting strips. The vortex-vortex interactions
and the edge barrier are calculated in the frame work of the screened Coulomb
model \cite{Minhagen}. With a change of scale the problem may be mapped into
the well known problem of an infinite slab under a parallel field. The
calculated $m(H)$ presents a series of maxima corresponding to matching of 
the VL in the pinning structure and a strong assymetry of the upward (virgin) 
and downward brenches, which charactere the surface effects. We also
study the dynamics of the VL when a transport current is applied accross 
the strip. The strong current inhomogenity leads to different dynamics in
different regions of the strip.

\section{Screened Coulomb model and dynamics}

We describe the motion of vortices in a superconducting strip of thickness 
$d$ much smaller than the London penetration depth $\lambda$. The geometry
and coordinate system are depicted in Fig. \ref{fig1}. The vortex
equation of motion, appropriate for overdamped dynamics, is $\eta{\bf v}_k 
= {\bf F}_L({\bf r}_k) - \nabla U_p({\bf r}_k) + {\bf\Gamma}_k(t)$, where
${\bf{v}}_k$ is the instantaneous terminal velocity of vortex $k$,
${\bf F}_L({\bf r}_k)$ is the total Lorentz Force, $U_{p}$ is the pinning
potential and ${\bf\Gamma}_k(t)$ is a Gaussian Langevin force acting on
vortex $k$ at instant $t$. The Lorentz force acting on a vortex line is
defined by the total sheet current ${\bf J} = \int{\bf j}dz$ probed by
this vortex, ${\bf F}_L = \phi_0{\bf J}\times\hat{z}$. A difficult task is
to precisely determine an expression for ${\bf F}_L$ in this system. Recently,
Brandt has generalized the continuum theory to account for the
penetration depth and numerically calculated the stream lines of current
of vortices in thin films and washers in the small $\Lambda$ limit
\cite{Brandt01}.

A usefull procedure is to  write the sheet current in terms of a scalar
potential, the stream function $g(x,y)$ satisfying the zero divergence of the
current \cite{Brandt92}: ${\bf J}(x,y) = \nabla\times \hat{z}g = -\hat{z}
\times\nabla g$. Note that there is a close relation between the stream 
lines probed by a vortex and the energy of this vortex \cite{Brandt01}.
This is clear when we rewrite the Lorentz force as ${\bf F}_L =
\phi_0{\bf J}\times\hat{z} = -\phi_0\nabla g$. With the help of the
Biot-Savart law, the London equation for a vortex located at ${\bf r}_k$
in the $xy$ plane of a thin strip under a perpendicular field $H$ may be
written as
\begin {equation}
  \Lambda\nabla^2 g({\bf r})-\! \int \! d^2r' K({\bf r,r'})
  g({\bf r'}) = H - \frac{\phi_0}{\mu_0}\delta({\bf r}-{\bf r}_k),
  \label{London}
\end {equation}

\noindent where $K({\bf r,r'})$ is a kernel with a suitable cutoff to account
for the finite sample thickness \cite{Brandt01}. We propose a simple way of
calculating $g$, using a screened Coulomb model approach, suitable for 
describing thin-film vortices with $\Lambda$ comparable to the system 
dimension \cite{Minhagen}. In this large-$\Lambda$ limit, the screening 
length $\Lambda_c$ is identified with the smaller of $\Lambda$ and some linear
dimension of the sample. Therefore, for thin strips with $W>\Lambda$, we use
a nonhomogeneous Helmholtz equation instead of solving the exact,
although much more complicated, Eq. \ref{London}.
Using the boundary condition $g=0$ at the edges, this problem is similar to
that of a superconducting slab under a parallel field \cite{Clecio01}.
Here, we add periodic boundary conditions (with periodicity $L$) in the
$y$ direction. Solving for a single vortex, one obtains the vortex-vortex
interaction energy:
\begin{figure}[t]
\centerline{\includegraphics[scale=0.325]{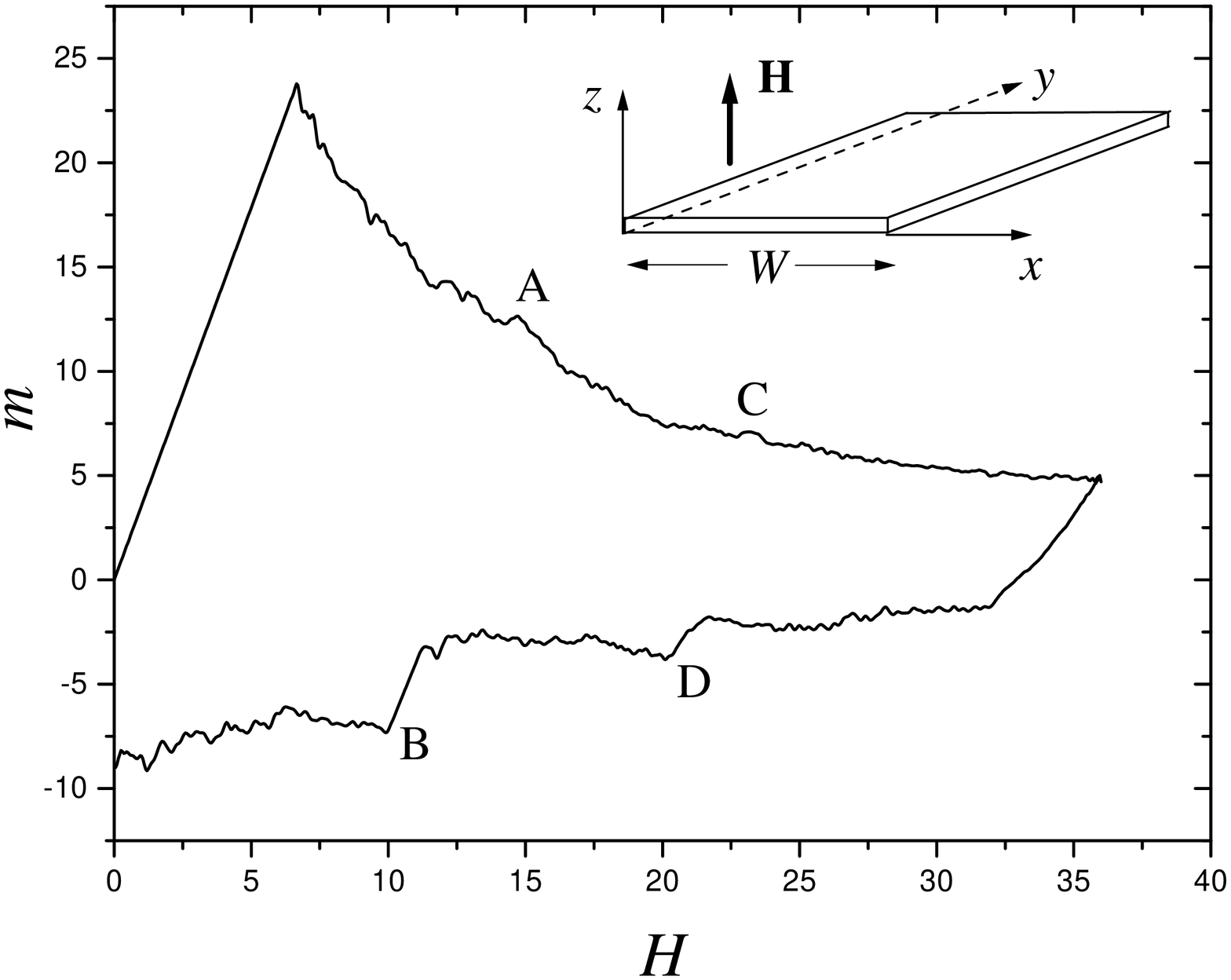}}
\vspace{-10mm}
\caption{As calculated magnetic moment $m$ vs. magnetic field $H$ for the thin 
strip with pinning strength $U_0 = 0.5\epsilon$. $m$ is in units of $\phi_0$ 
and $H$ in units of $\phi_0/\mu_0\Lambda_c^2$. Inset: system geometry and
coordinate system.}
\label{fig1}
\end{figure}
\begin{eqnarray}
V_{vv} = \epsilon\sum^{\infty}_{m=1}
\frac{\cosh\alpha_{m}(\vert{\tilde{y}_i-\tilde{y}_j\vert}-L/2)}
{\alpha_{m}\sinh\alpha_{m}L/2}\times \nonumber \\
\sin\frac{m\pi\tilde{x}_i}{\widetilde W}\,
\sin\frac{m\pi\tilde{x}_j}{\widetilde W},
\label{Vvv}
\end{eqnarray}
\noindent where $\alpha_m =\sqrt{1+(m\pi/\widetilde{W})^2}$,
$\epsilon=\phi_0^2/\mu_0\Lambda_c W$.
The tilded length quantities are normalized by $\Lambda_c$. With a suitable
cutoff to account for the vortex core, one obtains the vortex self-energy:
\begin{eqnarray}
V_{\rm self} = \epsilon\sum^{\infty}_{m=1}
\frac{\cosh\alpha_{m}(\tilde{\xi}-L/2)}{\alpha_{m}\sinh\alpha_{m}L/2}
\,\sin^{2}\frac{m\pi\tilde{x_i}}{\widetilde W},
\label{Vself}
\end {eqnarray}
where $\xi$ is the coherence length. Now, solving for the external field
$H$, we have
\begin{eqnarray}
V_H = \phi_0 \frac{H}{\Lambda_c}\bigg[\frac{\cosh(\tilde x_i -
 \widetilde{W}/2)}{\cosh \widetilde{W}/2}-1\bigg].
\label{VH}
\end{eqnarray}
Equations (\ref{Vself}) and (\ref{VH}) form the edge barrier for flux
entry. Thus, the Lorentz force on a vortex $i$ due to a vortex $j$ and an
external field $H$, for instance, is given by $-\nabla_i V({\bf r}_i,
{\bf r}_j)$ where $V({\bf r}_i,{\bf r}_j) = V_{vv}({\bf r}_i,{\bf r}_j) +
V_{\rm self}({\bf r}_i) + V_H({\bf r}_i)$.

\section{Periodic pinning: static aspects}

In the above framework, we simulated $m(H)$ measurements on a thin strip
with $W = 3\Lambda$. A square lattice of columnar defects (SLCD) is
represented by a periodic arrangement of Gaussian wells with radius
$r_p = 0.045\Lambda$ and period $a_p = 0.3\Lambda$. The depth of the wells
is varied from $U_0 = 0$ to $U_0=\epsilon$. The simulations are
carried out using the same procedure as in Ref. \cite{Clecio01}: at each
time step a vortex is attempted to nucleate in some random point in a line
which is a distance $\xi$ from one of the strip edges. Here we take
$\widetilde{\xi} = 0.01$. The magnetic moment was calculated using
${\bf m} = \frac{1}{2}\int d^2r\:{\bf r}\times{\bf J} = \frac{1}{2}\int d^2r
\:{\bf r}\cdot\nabla g$, which for the chosen gauge of $g=0$ at the edges is
simply $\int d^2r\,g$.
\begin{figure}[t]
\centerline{\includegraphics[width=19pc,height=18pc]{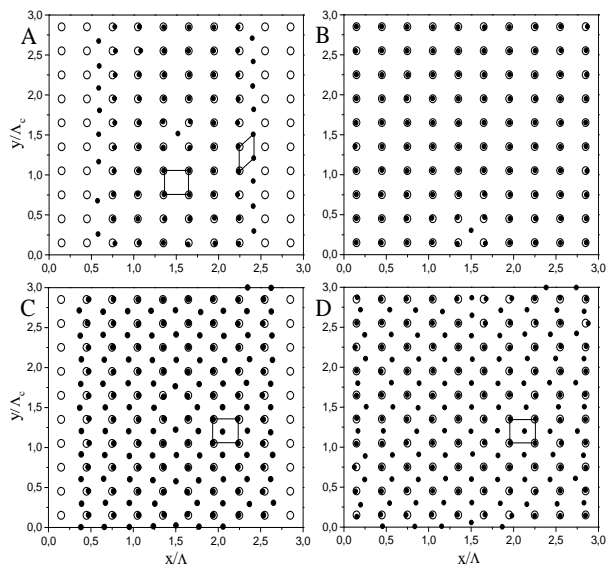}}
\vspace{-8mm}
\caption{Snapshots of the vortex lattice (closed circles) in the matching
peaks depicted in Fig. \ref{fig1}: (A) $H=14.5$ (upward branch), (B)
$H=10.5$ (downward), (C) $H=23$ (upward), and (D) $H=20.5$ (downward). The
hollow circles represent the pinning lattice.}
\label{fig2}
\end{figure}

In Fig. \ref{fig1} is shown the first $m(H)$ half cycle for $U_0 =
0.5\epsilon$. This curve depicts features of both bulk pinning due to 
regularly arranged deffects and edge effects. Matching peaks are present, 
but behave differently in the upward (virgin) and downward branches. As shown
in Fig. \ref{fig2}, these peaks correspond to hystory dependent states of 
commensuration. In upward field, the edge barrier imposes a vortex free 
region which produces vacant pinning sites even in the second matching 
configuration. In downward field, these states are similar to those expected
for infinite samples, no vortex free region is present. This points up the 
different dynamics of flux entrance and exit.

\section{Periodic pinning: dynamical aspects}

We have also simulated measurements of the time averaged CM velocity of the
VL as a transport sheet current $J_t$ was varied. The current is apllied
along the $y$ direction and the contacts are considered to be far away from
the region studied, in such a way to guarantee zero current divergence in
this region. The current distribution in the strip width may be calculated
from the Helmholtz equation for $g_J$ with the condition that $g_J$ must be
an odd function and that $g_J(W) - g_J(0) = J_t W$ (current
continuity). Thus the potential energy due to $J_t$ is
\begin{eqnarray}
V_J = \frac{J_t W\sinh(\tilde x_i - \widetilde{W}/2)}
{2\sinh\widetilde{W}/2} \label{VJ}
\end{eqnarray}
\begin{figure}[t]
\vspace{-5mm}
\centerline{\includegraphics[width=20pc,height=19pc]{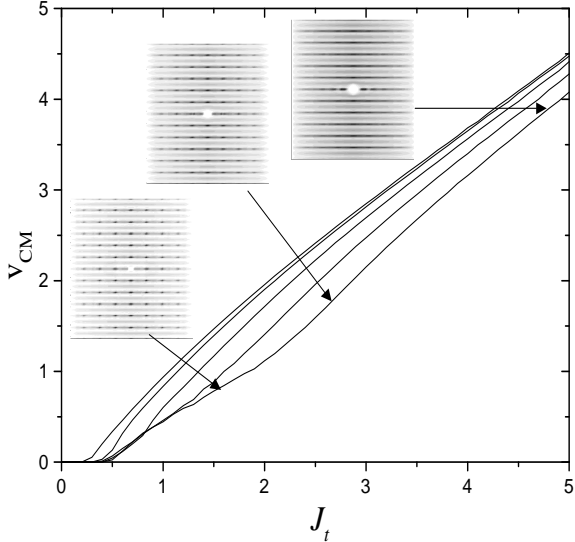}}
\vspace{-10mm}
\caption{Time-averaged CM velocity $v_{CM}(J_t)$ in the central third of the
strip for pinning strengths $U_0 = 0$,0 .25, 0.5, 0.75,
and 1.0$\epsilon$ (from top to botton). Panels: density-density correlation
functions for $U_0 = \epsilon$ and different current values (arrows).}
\label{fig3}
\end{figure}

The simulation procedure is the same as described in the preceeding section.
We divide the strip in three regions of width $\Lambda_c$. We computed the
time-averaged CM velocity $v_{CM}$ in each region as a function of the transport
current after the field has been ramped up to $24\phi_0/\mu_0\Lambda_c^2$.
In Fig. \ref{fig3} we show the current dependence of $v_{CM}$ in the central
region for different pinning strengths varying from $U_0 = 0$ to $U_0=\epsilon$.
As expected, there is a finite critical current for zero pinning, since the
edge barrier pins the VL. For nonzero pinning, two regimes of motion
are identified: in the low current regime the interstitial vortices are
depinned and start moving; at higher value of $J_t$ all the lattice is moving.
As shown by the density-density correlation function of the vortices position
(panels of Fig. \ref{fig3}), no reordering occur in the current
range studied. Reordering would be expected at least for
weak pinning since the temperature used in the simulations is a small fraction
of the zero-drive melting temperature. But here, lattice ordering is
contaminated by the random nature of flux penetration through the surface
barrier. However, long range time order was observed.

\section{Summary}

We have described the dynamics of vortices in narrrow superconducting 
strips in the screened Coulomb model approach. We calculated the
vortex-vortex interaction energy and the interaction with an edge barrier. 
In this scenario, we carried out simulations on vortex motion in field 
ramp measurements of the magnetic moment and I-V characteristics using 
Langevin dynamics. The $m(H)$ curves evidenced the strong influence of
the edge barrier in the VL configuration in upward field, where the first 
matching states are not fully accomplished in the entire sample. For the I-V
characteristics, we observed that the edge barrier represents an effective
pinning for the VL as expected, and ordered motion of the VL is disturbed
by the random penetration of vortices. A more detailed study will be reported
elsewhere.


This work was sponsored by the Brazilian agencies CNPq, FACEPE and 
FINEP.


\end{document}